# A Platform for All-optical Thomson/ Compton Scattering with Versatile Parameters

Siyu Chen,[1,2] Wenchao Yan,[1,2,a)] Mingyang Zhu,[1,2] Yaojun Li,[1,2] Xichen Hu,[1,2] Hao Xu,[1,2] Jie Feng,[1,2] Xulei Ge,[1,2,3] Wenzhao Wang,[1,2] Guangwei Lu,[1,2] Mingxuan Wei,[1,2] Lin Lu,[1,2] Xiaojun Huang,[1,2,3] Boyuan Li,[1,2] Xiaohui Yuan,[1,2] Feng Liu,[1,2] Min Chen,[1,2] Liming Chen,[1,2,b)] Jie Zhang[1,2,3]

**AFFILIATIONS:**

[1]*Key Laboratory for Laser Plasmas, School of Physics and Astronomy, Shanghai Jiao Tong University, Shanghai 200240, China.*
[2]*Collaborative Innovation Center of IFSA, Shanghai Jiao Tong University, Shanghai 200240, China。*
[3]*Tsung-Dao Lee Institute, Shanghai Jiao Tong University, Shanghai 201210, China.*

a)Author to whom correspondence should be addressed: wenchaoyan@sjtu.edu.cn
b)Author to whom correspondence should be addressed: lmchen@sjtu.edu.cn

**Abstract**:
A dual-beam platform for all-optical electron-photon scattering, or Thomson/Compton scattering, with adjustable collision-angle and parameter tuning ability has been developed, which, in principle, can be used for the verification of strong-field quantum electrodynamics effects. Combining this platform with a 200 TW Ti:Sapphire laser system, we demonstrated the generation of inverse Compton scattering X/gamma-rays with tunable energies from tens of keV to MeV. The polarization of X/gamma radiation was manipulated by controlling the polarization of scattering laser. In the near future, by combining this experimental platform with multi-PW laser facilities, it is proposed to experimentally generate X/gamma radiation with orbital angular momentum for the nuclear isomer excitation, and more importantly, to explore the regime transition from nonlinear Thomson scattering to nonlinear Compton scattering, eventually to demonstrate the verification of theories on extremely strong field quantum electrodynamics effects.

## I. INTRODUCTION

The investigation of strong-field quantum electrodynamics (SF-QED) processes requires extreme field intensities, approaching the Schwinger limit [1], which cannot be reached by state-of-art laser facilities. However, the existing laser intensity, when Lorentz-transformed in the rest frame of a relativistic electron during collision, can reach such high levels. Typically, when a laser photon undergoes an elastic collision with an electron, the frequency of the scattered photon remains consistent with the incident photon. Concurrently, the energy of the electron remains invariant before and after the scattering. This process is known as Thomson scattering [2]. However, when the laser intensity is increased, the interaction process induces an inelastic collision between the photon and the electron, this phenomenon referred to as Compton scattering [3]. In this scenario, the variation in the electron energy becomes non-negligible. Therefore, the Compton scattering process can be used to probe the SF-QED. In the laboratory frame, this process typically involves electrons transferring energy to scattered photons, hence it is commonly referred to as inverse

Compton scattering (ICS). However, in the rest frame of the electron, where the interaction process itself is often under investigation, it continues to be termed as Thomson scattering.

To systematically study the electron-photon interaction process, we have developed an experimental platform for dual-beam all-optical Thomson/Compton scattering (EPATCS) with versatile parameter tuning. The EPATCS is an all-optical tabletop electron-photon collision setup, where a high-quality, high-energy electron beam is generated through laser wakefield acceleration (LWFA) by one of the laser beams [4], and the second laser beam is employed to interact with the relativistic electron beam. This all-optical experimental setup leverages the fundamental properties of optics to enable higher precision in electron-laser collision timing and spatial resolution. Through the manipulation of the intensity, polarization state, and field mode of the collision laser, the scattering electron beam emits X/gamma-rays with varying characteristics, as well as a high-quality strong radiation source for applications such as high-density dynamic imaging [5–12], photonuclear physics [13,14], investigation of radiation reaction (RR) [15,16]. Increasing the intensity of both laser beams is beneficial to further explore SF-QED processes, such as angular momentum transfer in the collision process and electron-positron pair production [17–19], to prove existing theories in the future.

Over the past decade, there has been a growing interest in all-optical electron-photon scattering [20,21], spurred by the advancement of super-intense ultrafast laser technology [22]. The principle of scattering is illustrated in Fig.1, where $\theta$ represents the collision angle between the laser and electron beams, $\gamma_e$ represents the Lorentz factor, $\omega_0$ represents the frequency of the scattering laser photon, and $\omega_{sc}$ represents the frequency of the emitted photon. There are two possible experimental configurations for these all-optical scatterings. The concise configuration involves a single laser beam that induces the wakefield within the plasma to accelerate electrons [23–29], while a plasma mirror is placed in the path to reflect the laser beam, subsequently colliding with the relativistic electrons and emit foreword X-rays. The other scheme features a layout with two independently adjustable optical paths [15,16,25,30–35]. Precise temporal and spatial synchronization between the femtosecond laser and relativistic electrons are necessary for this scenario. The dual-beam scheme offers the distinct advantage of enabling control over the properties of the generated X-ray radiation by independently altering the properties of the scattering laser. Therefore, the radiation source with the dual beam experimental layout has more practical application and development prospects.

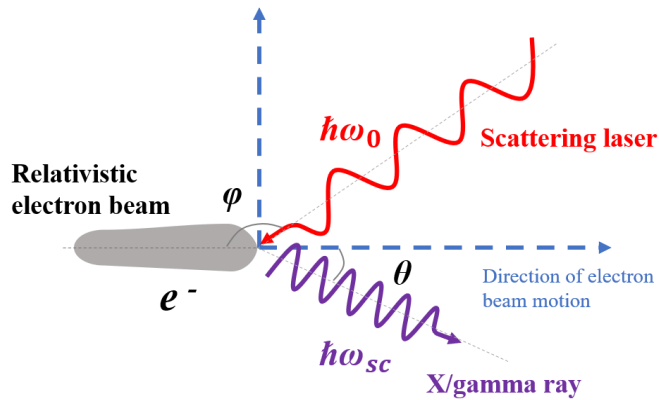

FIG. 1. Schematic diagram of Thomson/Compton scattering process. A laser-plasma accelerated electron beam is generated. Electrons oscillate transversally and emit X/gamma-rays due to this process.

The dual-beam layout offers distinct advantages. Each of the two lasers can be individually tuned to optimal conditions, enabling the generation of high-energy electron beams and the scattering laser that meet specific requirements. The quasi-monoenergetic femtosecond electron beams with diverse energies generated by LWFA can be obtained by varying the injection mode and the effective acceleration length. With an increase in the intensity of the scattering laser, the scattering mechanism gradually change from elastic to inelastic scattering in the electron frame, and the electrons oscillate and decelerate in the strong laser electromagnetic field. The interaction process involves a progression from linear to nonlinear dynamics, culminating in SF-QED, corresponding to typical effects and the generation of various forms of matter.

The products of such interactions can serve as high-quality compact ICS radiation sources. Previous series of experiments were often limited to adjusting parameters such as electron energy or laser intensity. The EPATCS can independently change multiple parameters, allowing for more precise adjustment of various radiation parameters. The radiation spectrum range would be significantly modified by transforming through the manipulation of the collision angle along the observation angle of electron movement direction. Consequently, the ICS based X/gamma-ray source spectrum cover tens of keV to MeV, tunable across a broad range. In addition, altering the polarization state of the scattering laser induces a corresponding transformation in the polarization state of the resulting gamma radiation. The polarization state of the emitted radiation can be modulated by controlling the colliding laser to be linear or circular. Moreover, high-energy gamma photons with orbital angular momentum (OAM) can be generated by altering the optical field mode of the scattering laser, which is expected to open avenues for investigating the physics of multiphoton absorption in the linear to nonlinear processes of Thomson scattering.

The EPATCS could also be built on the Petawatt laser facilities, and such a dual-beam platform can lead to the collision of GeV high-energy electrons and laser intensity, reaches the SF-QED regime [36]. The platform would be dedicated to the study of theories under extreme conditions, including RR, pair production, angular momentum transfer and conversion in the collision process, etc.

This paper is organized as follows: In Sec. II we present the basic parameter of the electron beam and X/gamma-ray on the platform. The Sec. III presents the experimental results based on the EPATCS at the 200 TW laser in Shanghai Jiao Tong University. In Sec. IV elucidate the prospective research of EPATCS.

## II. BASIC PARAMETERS OF THE ELECTRON BEAM AND X/GAMMA-RAY SOURCES

### A. Relativistic electron beam

The acceleration gradients based on LWFA can reach up to 1 GeV/cm, significantly exceeding those of conventional radiofrequency technologies by three orders of magnitude. Once the electrons are injected into the cavity structure, they undergo rapid acceleration driven by the intense electromagnetic field. After traversing the effective acceleration length, the electron acquires a relativistic velocity with a divergence angle on the order of milliradians in vacuum. A stable plasma wakefield is produced when the size of the laser spot matches the density of the plasma generated by the nozzle [37–39]. For stable acceleration of the injected electrons with different injection mechanisms [40–47], it is crucial that the cavity structure formed by the wakefield is large enough to produce a steep acceleration gradient and a narrow sheath of the focusing field at the tail of the channel. High-quality and high-energy ultrafast electron beams can be obtained when electrons are accelerated by a stable wakefield. The maximum energy of the electron beam is related to plasma density ($n_p$) as follows [39]:

$$\Delta E\,[\text{GeV}] \simeq 1.7 \left(\frac{P\,[\text{TW}]}{100}\right)^{1/3} \left(\frac{10^{18}}{n_p\,[\text{cm}^{-3}]}\right)^{2/3} \left(\frac{0.8}{\lambda_0\,[\mu\text{m}]}\right)^{4/3}, \quad (1)$$

where $P$ donates the laser power at the target and $\lambda_0$ donates the laser wavelength.

In these experiments, we focused a 200 TW femtosecond laser, with the wavelength centered at 805 nm, to a focal spot of 20 μm (FWHM) using a long focal length off-axis parabolic mirror (F#20). A supersonic gas nozzle was used to produce the under-dense gas target. By extracting a portion of the laser from the main optical path and focusing it above the nozzle after passing through a delay system, the electron beams were trapped via optical injection, then accelerated. The electron beams with peak energies ranging from 50 MeV to 300 MeV were obtained by varying the position of injecting laser beam relative to the nozzle and then altering the acceleration length. The typical electron beam profiles, indicating the electron energy as shown in Fig.2, were recorded by a phosphor screen after the electron spectrometer with a 1-T, 16-cm-long magnet. In addition, the consecutive electron energy spectra are conducted, within the electron beam divergence angle is stable in the range of 4 ± 0.5 mrad under constant conditions of gas density and laser.

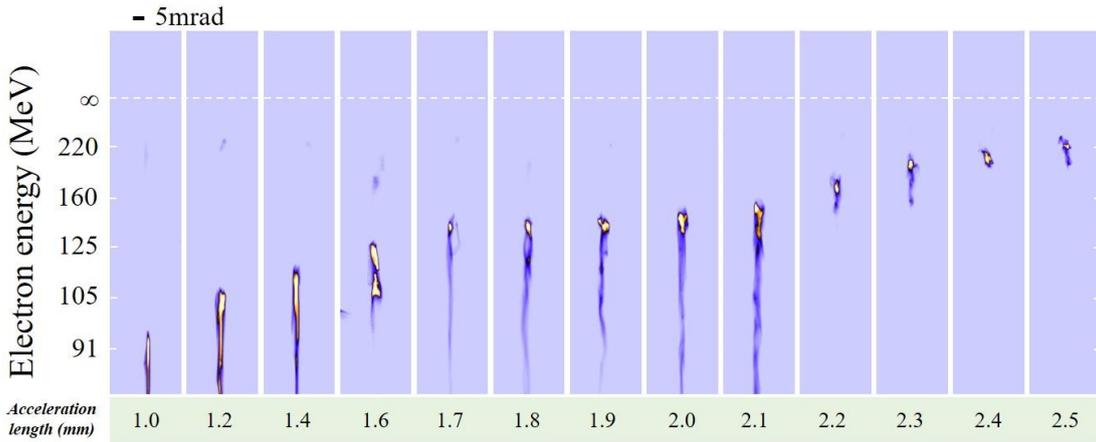

FIG. 2. The diagnosis results of the electron beam with different acceleration lengths by optical injection.

## B. X/gamma-ray in Inverse Compton Scattering (ICS)

When the relativistic electron beams collide with the scattering laser beam at the focal position, they oscillate under the strong laser electric field and generate forward radiation. In the laboratory frame, the photon frequency experiences Doppler shift twice, resulting in the radiated frequency $\omega_\gamma = 2\gamma_e \omega_0' = 4\gamma_e^2 \omega_0$. This represents the fundamental relation between the frequencies of incident and scattered photons in the laboratory frame.

The intensity of the laser electric field that accelerates the electron from rest energy to the speed of light in one laser period is defined as the normalized electric field intensity, $a_0 = eA_0/m_e c \simeq 0.85 \times \lambda[\mu m]\sqrt{I_0[10^{18}\,\text{W/cm}^2]}$, where $A_0$ represents the amplitude of the laser electric field, $e$ denotes the electron charge, $m_e$ represents the electron rest mass, and $c$ represents the speed of light. When $a_0 \ll 1$, the relationship between the outgoing photon frequency and the scattering laser photon frequency is described as above. In this regime, the electron deceleration and photon recoil effect are usually ignored, and the nonlinear effect is a perturbation parameter, thus the process is referred to as linear Thomson scattering. When $a_0 \gtrsim 1$, the Lorentz force induced by the magnetic field become pronounced, leading to more complex electron motion. As a consequence, the electron absorbs multiple laser photons and emits high-energy gamma photons, and the radiation spectrum would contain high-order harmonics, known as the nonlinear Thomson scattering. As the intensity of the scattering laser increases, the linear process of single-photon absorption gradually changes to a nonlinear process of multiphoton absorption. Therefore, the exact expression for the nonlinear process is as follows [48]:

$$E_{sc} = E_0 \frac{2n\gamma_e^2(1-\cos\varphi)}{1+a_0^2/2+\gamma_e^2\theta^2}, \qquad (2)$$

where $n$ represents the nonlinear order, $E_0$ represents the incident photon energy, $\gamma_e$ represents the relativistic factor of the electron, $\varphi$ represents the angle between the electron beam and the scattering laser, and $\theta$ represents the radiation observation angle.

In the linear head-on scattering regime, the energy of the scattered photon is directly determined by the energy of the electron. As the energy of the electron varies by one order of magnitude, the energy of the forward radiation varies by two orders of magnitude. Within a specified range of the colliding laser intensity $a_0$, the fundamental frequency radiation redshifts and the energy decreases. The spectral width of the radiation is determined by the energy dispersion of the relativistic electron beam and $a_0$. When the nonlinear effect becomes prominent, the X/gamma-ray spectrum significantly broadens. The duration of radiation scattering primarily depends on the interaction time between the relativistic electron beam and the scattering laser.

### III. VERSATILE PARAMETER TUNING IN EPATCS

The enhancement of ICS flexibility not only bolsters the robustness of the experimental platform, but also allows for customization based on diverse scientific objectives. The all-optical inverse Compton scattering (AOICS) is regarded as a high-quality X/gamma radiation source. Therefore, the high-precision and efficient modification of the radiation spectrum range, along with the free adjustment of the radiation polarization state, lays a crucial foundation for the development of EPATCS.

**A. Tunable X/gamma-ray energy**

According to Eq. (2), the energy of scattered photons is correlated with that of the incident photons in the following manner, $E_{sc} \sim E_0 \gamma_e^2 \eta$, where $\eta = 1 - cos\varphi$ is a parameter related to the collision angle, exhibiting a clear range from 0 to 1. Therefore, to make the outgoing radiation energy range of AOICS continuously adjustable, altering the collision angle is an effective method. Fig. 3(a) shows the dual-beam layout with flexible collision angles, which can achieve an energy-tunable AOICS process. As shown in Fig. 3 (b), under a given electron beam energy, the radiation energy of observation angle $\theta = 0$ changes with the collision angle $\varphi$.

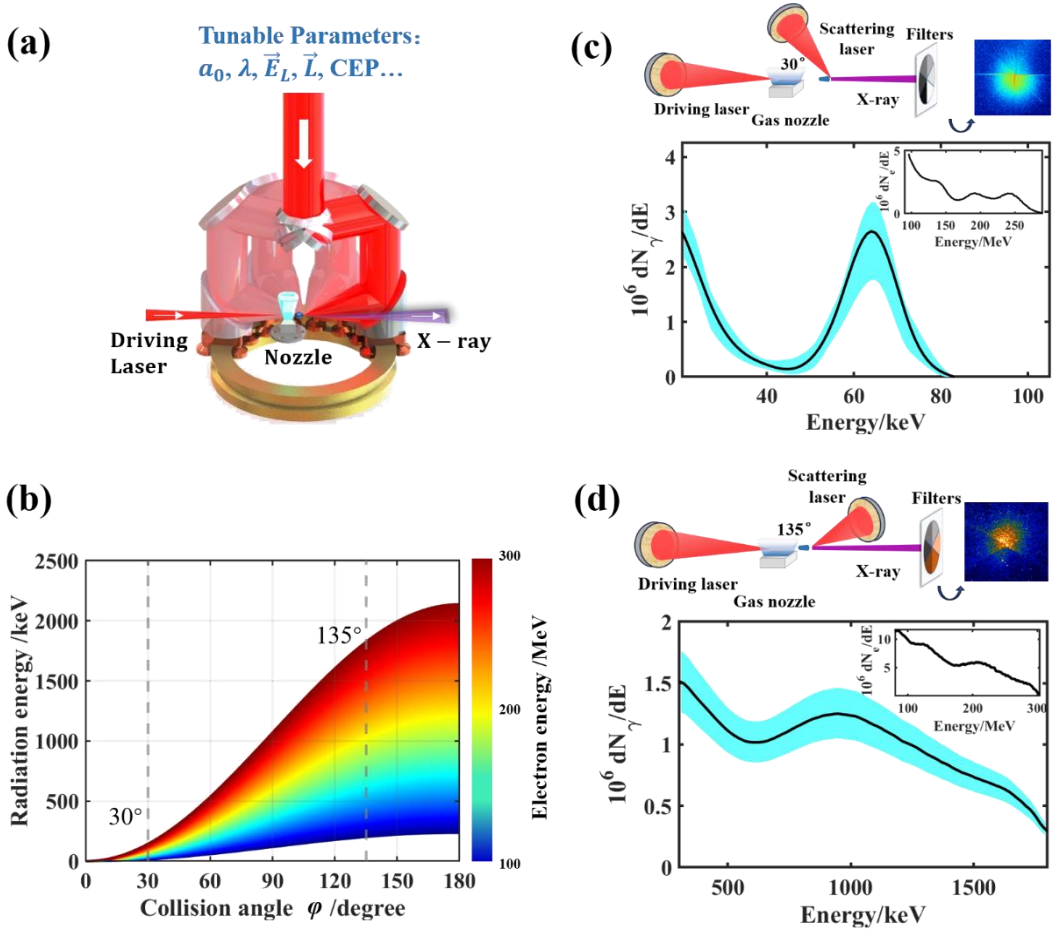

FIG. 3. (a) The schematic diagram of experimental layout with multiple collision angle. (c) and (d) are the experimental layout diagrams of AOICS under two conditions of 30° and 135° collision angles, and the radiation spectra measured by the two angles, and the electron energy spectrum results for that time are shown in the upper right corner of each graph, respectively. (b) represents the radiation energy under different collision angle $\varphi$ with electron energy from 100 MeV to 300 MeV when the observed angle $\theta = 0$.

In the experiment, the relativistic electrons generated by LWFA has a cutoff energy of approximately 300 MeV, approaching a continuous spectrum. Two experimental configurations of collision angle $\varphi = 30°$ and $\varphi = 135°$ respectively are set, while the intensity $a_0$ of the collision laser is ~ 1. The range of the radiation spectrum can be determined by altering the collision angle using Eq. (2). Metal filter sheets of various materials and thicknesses are designed for radiation diagnosis. For the experimental layout with a 30° collision angle, the filter combination is based on the K-absorption edge of metal filters of different materials and thicknesses. Through the least squares method, we obtained convergence results and carried out the energy spectrum. The resulting error stems from the width of the K-absorption edge of the filter and the response of the image plate. The quasi-monoenergetic peak at about 65 keV in the results may originate from ICS colliding with only a fraction of electrons. In the case of a 135° collision angle, the radiation energy spectrum spans higher than 1 MeV. While there is no suitable metal atom K-absorption edge for reference, matching filters are selected, ensuring that the X-ray transmission of the adjacent filter is different and covered as much as possible. The iterative least square method for numerical analysis of transmission coefficient is adopted [49]. Thus, the final energy spectrum results are obtained, as illustrated in Fig. 3 (c) and (d). The radiation spectra at two collision angles correspond with the electron energy spectrum. Due to the relatively lower diagnostic accuracy of the high-energy radiation spectrum, a wider radiation energy spectrum range is obtained when the collision angle is larger.

LWFA is capable of generating high-energy electrons that can reach multi-GeV with a suitable length nozzle. Furthermore, it is anticipated that the first-order energy spectrum of gamma radiation will extend to cover hundreds of MeV, paving the way for broader applications in the future.

**B. Polarization control of X/gamma ray**

Polarization is a fundamental characteristic of electromagnetic radiation [50]. Polarized X-rays possess the unique ability to probe the characteristics of magnetic structures in structural magnetism and distinguish between chiral and helical magnetic structures [51–55]. Moreover, gamma ray polarimetry in AOICS opens new perspectives for nuclear resonant scattering experiments [56–59]. In the AOICS process, an accurate diagnosis of the resulting incoherently polarized X-rays is particularly crucial. Extreme sensitivity polarization diagnostics enable the study of nonlinear properties of the vacuum. In the QED description of vacuum, virtual particle-antiparticle pairs, known as quantum fluctuations, are allowed to exist for ultra-short periods of time [60,61]. In the presence of a strong external electric or magnetic field, these virtual particle pairs can partially align, resulting in the optically polarized nature of the vacuum.

Based on the EPATCS, we generate polarized X-rays via the collision of a polarized laser beam and the relativistic electron beam. In the scattering laser path, a half-wave plate or a quarter-wave plate is incorporated to change the polarization state of the scattering laser, thereby modifying polarization state of the generated radiation accordingly. An off-axis parabolic mirror of F#5 is used to generate and diagnose linear and circular gamma rays via AOICS. The experimental layout is illustrated in Fig. 4 (a).

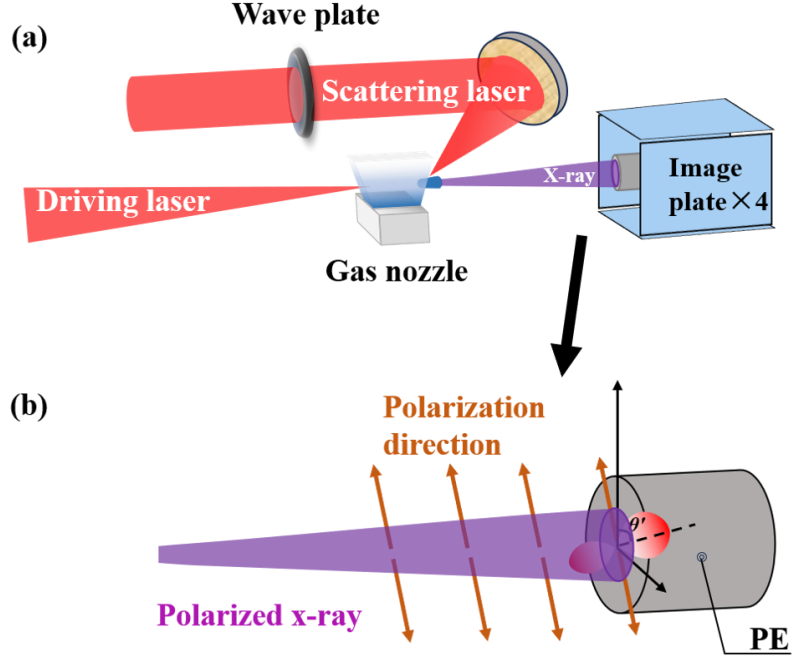

FIG. 4. (a) Experimental layout. The polarization state of the X-ray was obtained by placing the polyethylene forward in the X-ray and placing four image plate around it to diagnose the signal scattered by the X-ray Compton in different polarization states. (b) Schematic representation of Compton scattering of linearly polarized X-rays with polyethylene scatterers.

As shown in Fig. 4 (a), the relativistic electron beam is deflected by a magnet. A cylindrical polyethylene converter with a diameter of 2 cm and a length of 15 cm is placed 1.9 m away from the impact point, and four image plates are positioned around the converter. The secondary photon signal radiated on the scatterer is diagnosed based on the Compton scattering. When the scattering angle is close to 90°, the azimuth distribution of scattered photons is heavily dependent on X-ray polarization, making Compton scattering effective for polarization analysis [62]. The expressions of the scattering cross section of Compton scattering in the vertical direction for linear polarization [see Eq. (3)] and circular polarization [see Eq. (4)] are as follows:

$$\frac{d\sigma_{lin\,\perp}}{d\Omega} = \frac{1}{4}r_e^2\left(\frac{\varepsilon}{\varepsilon_0}\right)^2\left[\frac{\varepsilon}{\varepsilon_0} + \frac{\varepsilon_0}{\varepsilon} - 2\cos^2\theta'\right], \qquad (3)$$

$$\frac{d\sigma_{cir\,\perp}}{d\Omega} = \frac{1}{4}r_e^2\left(\frac{\varepsilon}{\varepsilon_0}\right)^2\left[\frac{\varepsilon}{\varepsilon_0} + \frac{\varepsilon_0}{\varepsilon}\right], \qquad (4)$$

where $r_e$ represents the classical radius of the electron, where $\varepsilon_0$ is the energy of incident photon, $\varepsilon$ the energy of the scattered photon, and $\theta'$ denotes the scattering azimuth angle. According to Eq. 3, the Compton scattering of linearly polarized X-rays through the scatterer in the vertical direction is a function of the azimuth angle $\theta'$. When the incident X-ray is circularly polarized, Eq. 4 reveals that the vertical Compton scattering is independent of the azimuth angle $\theta'$. In the experiment, the optical axis of the half-wave plate is adjusted to an angle of 22.5° from the original horizontal polarization direction, thus producing a linear polarization angle of 45° from the horizontal direction, as shown in Fig. 4 (b). The reason behind this choice of angle is to facilitate the differentiation of the background signal. In the case of circularly polarized scattering laser generated with a quarter wave plate, the actual transmitted laser is circularly polarized by about 80% due to the wide spectral width of the laser, which is affected by the bandwidth of the wave plate.

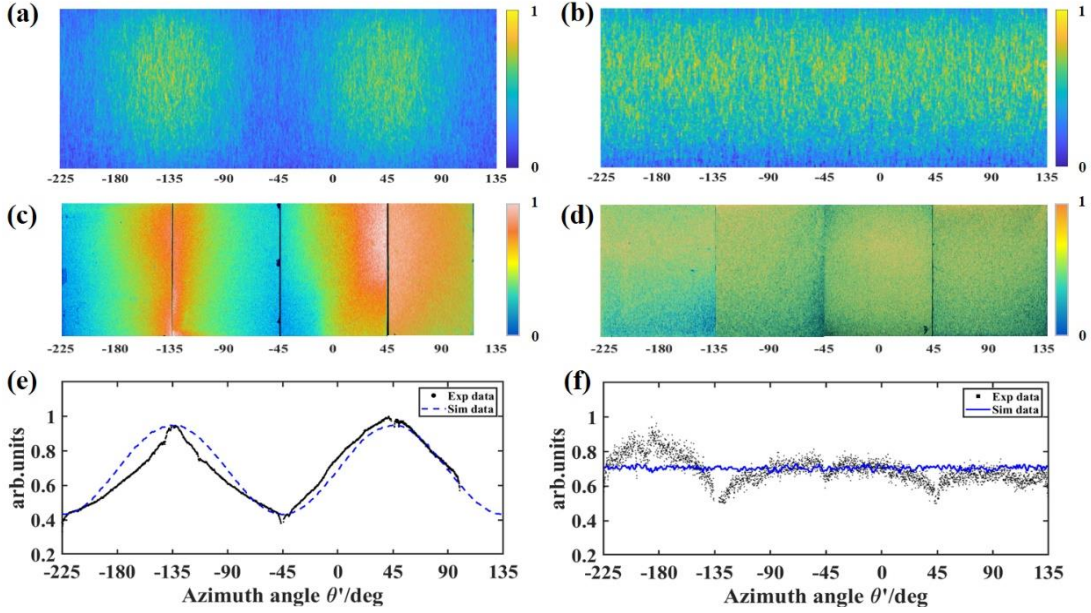

FIG. 5. (a), (b) show the simulation results by FLUKA, corresponding to the respective Compton scattering signals of linearly polarized and circularly polarized X-rays with polyethylene. (c), (d) is the experimental diagnostic result of linearly polarized and circularly polarized X-rays, respectively. In (e), (f), the signal image formed by black dots is the one-dimensional integral result of the experimental results, and the dashed blue line represents the simulation results.

We conducted simulations using FLUKA to simulate the process, and the results are shown in Fig. 5 (a) and (b). The experimental results are obtained, as shown in Fig. 5 (c) and (d), accumulated 100 times of linearly polarized and circularly polarized X-rays, respectively. The four image plates are arranged corresponding to azimuth angles ranging from -225° to 135°, with the portion corresponding to 100°-135° being absent, due to the constraints on the size of the image plate. One-dimensional integral correspondence is performed between the experimental and theoretical results, as shown in Fig. 5 (e) and (f). The linear polarization diagnostics show a periodic intensity distribution over the four IP, which is definitely a Compton scattering signal corresponding to linearly polarized X-rays and scatters. The degree of polarization $(I_{max} - I_{min})/(I_{max} + I_{min})$ according to experimental results is about 0.42, and the background greatly influenced such a result. Moreover, the diagnostic results of circularly polarized X-rays show uniformly distributed intensity signals overall, although there is still a weak periodic intensity distribution signal near 0° and 180°. This suggests that the polarization degree of circularly polarized X-rays generated by the AOICS is not 100%, but rather the elliptical polarization state along the transverse axis. Such a result matches the effect of the waveplate bandwidth described above. It is evident from the results that the X-ray diagnosis results of linear polarization and circular polarization exhibit their own incomplete symmetry in orientation, which arises due to the off-center position of the polyethylene scatterer incident by X-ray [29].

## IV. A ROADMAP FOR PROSPECTIVE RESEARCH

By integrating with multi-PW laser facilities, the adjustable parameters of the EPATCS lay vital groundwork for forthcoming research endeavors, encompassing multiphoton Thomson/Compton scattering, radiation reaction, vacuum polarization effects, and electron-positron pair production [36,63–67]. The spatial synchronization accuracy of EPATCS can, in principle, reach the spatiotemporal synchronization accuracy on the level of '$\lambda^3$' [68]. Such high spatiotemporal synchronization accuracy can precisely control the interaction position of the electron beam and the focused laser, which is conducive to investigate the SF-QED process, as well as generating high-quality unique radiation source, such as radiation with orbital angular momentum.

**A. Experimental plan on SF-QED**

Compton scattering involving high-energy electrons and super-intense lasers serves as a key experimental approach to verify the effects of SF-QED under extreme conditions. The quantum nonlinearity parameter is $\chi_0 = \sqrt{(F^{\mu\nu}p_\mu)^2}/(E_s m_e c)$, where $F^{\mu\nu} = \partial_\mu A_\nu - \partial_\nu A_\mu$ is the four-tensor EM field, $p_\mu$ is the particle four-momentum, and $E_s = m_e^2 c^3/e\hbar$ is the Schwinger limit field. When $\chi_0 \gtrsim 1$, the quantum effect of radiation damping can no longer be ignored, which is called the quantum radiation dominant regime (QRDR) [66,69,70]. In this case, both the quantum effect and the RR effect dramatically alter the dynamic behavior of the electrons. Under such conditions, RR can be fully explored, and the generation of electron-positron pairs can be realized.

Over the next few years, the EPATCS is planed to undergo a three-phase development process to delve into SF-QED through Compton scattering. The table below presents the fundamental experimental proposal and parameters.

TABLE. The fundamental experimental proposal and parameters.

| | Fundamental Experimental Proposal | Electron Energy $\gamma$ | Peak Laser Intensity $a_0$ | Quantum Parameter $\chi_0$ |
|---|---|---|---|---|
| Phase I | From linear to nonlinear regime of ICS; Precision experiment of RR. | ≲2000 | 20 | ≲0.2 |
| Phase II | Research studies of CRDR for Compton scattering. | ~4000 | 40~60 | ~1 |
| Phase III | Research studies of QRDR for Compton scattering and BW process. | ~20000 | ~200 | ~20 |

In phase I, the EPATCS will be transferred and built in multiple advanced laser facilities. The platform will be transferred to the 0.5 PW femtosecond laser facility, which is currently commissioning in Key Laboratory of Laser Plasma at Shanghai Jiao Tong University. According to the matching condition of laser wakefield acceleration, low energy dispersive high-energy electron beams with maximum energy less than 1 GeV can be generated stably. Another focused laser beam with radius of 2 μm in FWHM, which power density can reach $10^{21}$ W/cm$^2$, corresponds to a normalized intensity of $a_0 \sim 20$. Under this premise, the quantum parameter is $\chi_0 \sim 0.2$, the platform can act as the four-dimensional spatial-temporal diagnostics of electron beams before and after collisions to demonstrate RR in a single shot.

In phase II, the platform will be transferred to the PW-level laser facilities, for instance, on a 1PW laser facility at the Synergetic Extreme Condition User Facility (SECUF) of the Institute of Physics (IOP) in Chinese Academy of Sciences. It can contribute to research studies of classical radiation dominated regime (CRDR) with normalized intensity $a_0 \sim 40$ [71,72] via Compton scattering. It can also be transferred to a 2.5 PW 800 nm Ti: Sapphire femtosecond laser system in Tsung-Dao Lee Institute (TDLI), with normalized intensity $a_0$ can be up to 60. At the meanwhile, high-quality electron beam generation with a central energy of 2 GeV, corresponding to $\gamma \approx 4000$, can be achieved using a several-meter-long focal length OAP focused laser that interacts with the gas cell. At this point, the quantum parameter is $\chi_0 \sim 1.4$, which make it possible to scrutinize the RR process. As the phenomenon become more pronounced, it would deepen the understanding the fundamental principle of SF-QED.

In phase III, the EPATCS could be transferred to the 10 PW laser facilities, such as SULF [71], SEL [72], APOLLON [73], ELI [74,75] and EP-OPAL [76]. Consider that a stable electron beam energy of multi-GeV could be generated, the normalized intensity $a_0$ of scattering laser would be up to 200, when the laser beam is focused to the diffraction limit, and the quantum parameter will be up to ~ 20. Under such conditions, quantum effects will become consequential, which will be conducive to research the experimental phenomena of ICS in QRDR. Moreover, when the high-energy electron beam collides with the laser, the pair production can be effectively detected, and experimental evidence for the occurrence of the Breit-Wheeler (BW) [17] process could be obtained. Additionally, combining with the Laguerre-Gaussian (LG) laser beam, the EPATCS will contribute to investigating SF-QED phenomena with complex spacetime structure, validating and exploring the locally monochromatic approximation (LMA) [77] and the local-constant-field approximation (LCFA) [78–82]. This could not only promote the occurrence conditions of SF-QED physical processes, but also further improve the SF-QED theories and experimental verification. The relevant international experimental progress and proposals are summarized as shown In the Fig. 6.

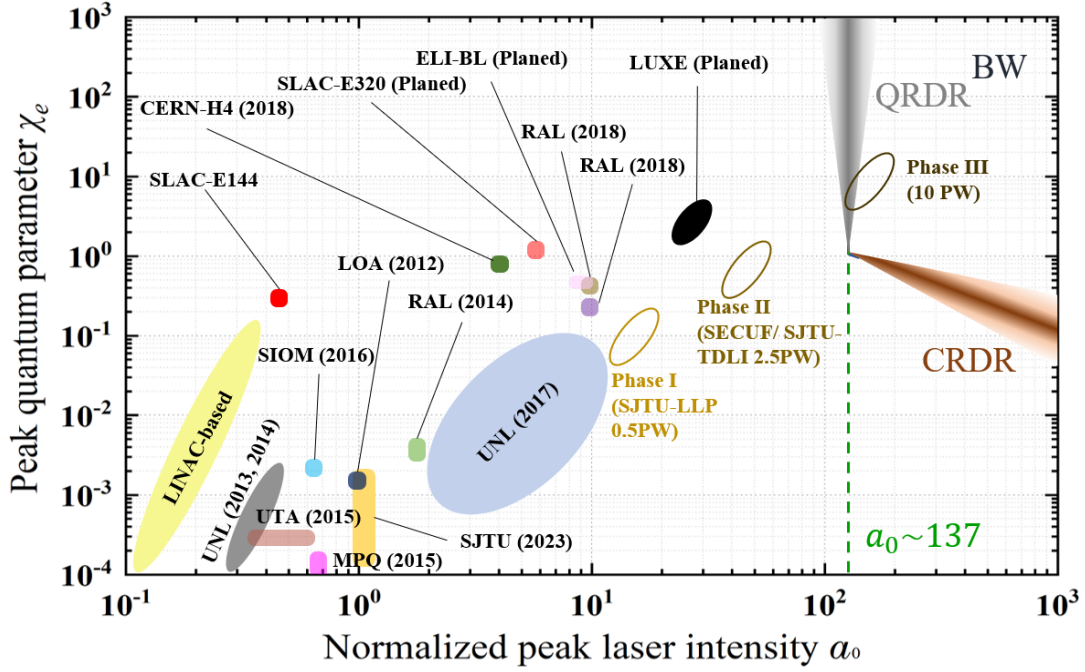

FIG. 6. The relevant international experimental progress and proposals.

**B. Development and application of the AOICS source**

The high-energy, ultrafast AOICS radiation has profound implications for the study of dynamic processes in dense matter [83]. The X/gamma-rays produced by the EPATCS, with photon energies ranging from tens of keV to MeV, offer a brightness at $10^{18}$-$10^{21}$ photons s$^{-1}$ mm$^{-1}$ mrad$^{-1}$ 0.1% BW, with a photon flux of about $10^7$-$10^{10}$ photons/shot. On the EPATCS, several plans will be carried out to improve the AOICS source, for example:

1) One of the novel approaches to enhance the photon flux of AOICS is to exploit the interaction between traveling waves and electrons. In 2010, A.D. Debus and M. Bussmann *et al*. [84] proposed the concept of "Traveling-wave Thomson scattering (TWTS)": the use of line-focused pulse front tilt to collide with electrons. The inclination angle of the pulse front of the laser is introduced by the grating, and then the pulse is focused on the direction line perpendicular to the paper plane line by using a cylindrical reflector whose focal line overlaps with the electron transmission path. It forms a microbeam and radiates a free-electron laser. This method can effectively improve the energy conversion efficiency of AOICS and increase radiated photon number.

2) The X/gamma-rays carrying orbital angular momentum (OAM) is generated through the AOICS with Laguerre-Gaussian (LG) mode of the scattering laser. This OAM radiation is a quantized representation, not a collective effect [85]. Our future work includes systematic experimental studies of OAM X/gamma-ray production, development of diagnostic methods for OAM gamma photons, and exploration of their application in nuclear science. The non-zero OAM of gamma photons could modify selection rules and generate higher multistage excited states [85,86], broadening our understanding of atomic nuclear energy states and opening new avenues in nuclear physics.

The AOICS source with versatile parameters and modes, will be developed for applications. By increasing photon energy, radiation luminosity, and stability, AOICS imaging will be enhanced, enabling imaging of ultrafast processes such as plasma generation or ablation. Furthermore, AOICS sources will be widely used in non-destructive testing, detection of weapon-grade materials, and customs inspections, nuclear physics.

## V. CONCLUSIONS

The development of an all-optical Thomson/Compton scattering platform, featuring versatile parameter tunability, harnesses the exceptional benefits of laser wakefield acceleration and super-intense ultrafast laser. This platform provides precise control over AOICS parameters, including the energy spectrum, polarization state, photon orbital angular momentum, and other parameters of high-energy radiation. With the ongoing development of super-intense ultrafast laser facilities, we are poised to conduct experimental studies from multiple perspectives, such as the experimental verification of SF-QED and the enhancement of the high-energy radiation generation and its applications.

## ACKNOWLEDGMENTS

This work is supported by the National Key R&D Program of China (2021YFA1601700), the National Natural Science Foundation of China (12074251, 11991073, 12335016, 12105174) and the Strategic Priority Research Program of the Chinese Academy of Sciences (Grant No. XDA25000000 (Grant No. XDA25030400)). We thank the sponsorship from Yangyang Development Fund.

## AUTHOR DECLARATIONS

**Conflict of Interest**

The authors have no conflicts to disclose.

**Author Contributions**

**Siyu Chen**: Formal analysis (lead); Writing – original draft (lead); Data curation (lead). **Wenchao Yan**: Conceptualization (lead); Methodology (lead); Writing – review & editing (equal). Project administration (lead); Resources (lead); Supervision (lead). **Mingyang Zhu**: Data curation (equal); Software (equal); Formal analysis (equal). **Yaojun Li**: Data curation (equal); Investigation (equal). **Xichen Hu**: Data curation (equal); Investigation (equal) ; Formal analysis (equal). **Hao Xu**: Data curation (equal); Formal analysis (equal). **Jie Feng**: Formal analysis (equal); Validation (equal). **Xulei Ge**: Software (equal); Validation (equal). **Wenzhao Wang**: Data curation (equal); Investigation (equal). **Guangwei Lu**: Investigation (equal). **Mingxuan Wei**: Investigation (equal). **Lin Lu**: Supervision (equal). **Xiaojun Huang**: Validation (equal); Supervision (equal). **Boyuan Li**: Software (equal); Validation (equal). **Xiaohui Yuan**: Validation (equal); Supervision (equal). **Feng Liu**: Software (equal); Validation (equal). **Min Chen**: Validation (equal); Supervision (equal).

**Liming Chen**: Methodology (equal); Project administration (equal); Writing – review & editing (equal); Validation (equal); Supervision (lead ). **Jie Zhang**: Validation (equal); Supervision (equal); Writing – review & editing (equal).

## DATA AVAILABILITY

The data that support the findings of this study are available from the corresponding author upon reasonable request.